\documentclass{aa}  

\usepackage{graphicx}

\usepackage{txfonts}
\usepackage[colorlinks=true, allcolors=blue]{hyperref}
\usepackage{orcidlink}
\usepackage{xspace}

\newcommand {\nustar}{{NuSTAR}\xspace}
\newcommand {\chandra}{{Chandra}\xspace}
\newcommand {\ixpe}{{IXPE}\xspace}
\newcommand {\xmm}{{XMM--Newton}\xspace}

\newcommand {\nicer}{{NICER}}
\newcommand\srga{\mbox{SRGA J1444}59.2$-$604207}

\newcommand{\fluxcgs}{erg\,s$^{-1}$\,cm$^{-2}$}
\newcommand{\lumcgs}{erg\,s$^{-1}$}

\graphicspath{{./}{figures/}}

\begin{document} 

\title{Discovery of polarized X-ray emission from the accreting millisecond pulsar {\srga}}

\titlerunning{Polarized X-rays from an accreting ms pulsar}

\authorrunning{Papitto et al.}
\author{
Alessandro Papitto \inst{\ref{in:OAR}}\thanks{        alessandro.papitto@inaf.it}\orcidlink{0000-0001-6289-7413}
\and Alessandro Di Marco \inst{\ref{in:INAF-IAPS}}\orcidlink{0000-0003-0331-3259}
\and Juri Poutanen \inst{\ref{in:UTU},\ref{in:RAS}}\orcidlink{0000-0002-0983-0049}
\and Tuomo Salmi \inst{\ref{in:UA},\ref{in:UH}}\orcidlink{0000-0001-6356-125X}
\and Giulia Illiano \inst{\ref{in:OAR},\ref{in:UniRoma2},\ref{in:LaSapienza}}\orcidlink{0000-0003-4795-7072}
\and Fabio~La~Monaca \inst{\ref{in:INAF-IAPS},\ref{in:UniRoma2},\ref{in:LaSapienza}} \orcidlink{0000-0001-8916-4156}
\and Filippo Ambrosino \inst{\ref{in:OAR}}\orcidlink{0000-0001-7915-996X}
\and Anna Bobrikova \inst{\ref{in:UTU}}\orcidlink{0009-0009-3183-9742}
\and Maria Cristina Baglio \inst{\ref{in:OAB}}\orcidlink{0000-0003-1285-4057}
\and Caterina Ballocco \inst{\ref{in:OAR},\ref{in:LaSapienza}}\orcidlink{0009-0001-0155-7455}
\and Luciano Burderi \inst{\ref{in:IASF},\ref{in:UCa}}\orcidlink{0000-0001-5458-891X}
\and Sergio Campana \inst{\ref{in:OAB}}\orcidlink{0000-0001-6278-1576}
\and Francesco Coti Zelati \inst{\ref{in:CSIC},\ref{in:IEEC}}\orcidlink{0000-0001-7611-1581}
\and Tiziana Di Salvo \inst{\ref{in:UPa}}\orcidlink{0000-0002-3220-6375}
\and Riccardo~La~Placa \inst{\ref{in:OAR}}\orcidlink{0000-0003-2810-2394}
\and Vladislav Loktev \inst{\ref{in:UTU}}\orcidlink{0000-0001-6894-871X}
\and Sinan Long \inst{\ref{in:UCL}}
\and Christian Malacaria \inst{\ref{in:OAR}}\orcidlink{0000-0002-0380-0041}
\and Arianna Miraval Zanon \inst{\ref{in:ASI}}\orcidlink{0000-0002-0943-4484}
\and Mason Ng \inst{\ref{in:MIT}}\orcidlink{0000-0002-0940-6563}
\and Maura~Pilia \inst{\ref{in:OAC}}\orcidlink{0000-0001-7397-8091}
\and Andrea Sanna \inst{\ref{in:UCa}}\orcidlink{0000-0002-0118-2649}
\and Luigi Stella \inst{\ref{in:OAR}}\orcidlink{0000-0002-0018-1687}
\and Tod Strohmayer\inst{\ref{in:Godd1},\ref{in:Godd2}}\orcidlink{0000-0001-7681-5845}
\and Silvia Zane \inst{\ref{in:UCL}}\orcidlink{0000-0001-5326-880X}
}
\institute{
INAF Osservatorio Astronomico di Roma, Via Frascati 33, 00078 Monte Porzio Catone (RM), Italy\label{in:OAR} 
\and 
INAF Istituto di Astrofisica e Planetologia Spaziali, Via del Fosso del Cavaliere 100, 00133 Roma, Italy \label{in:INAF-IAPS}
\and
Department of Physics and Astronomy, 20014 University of Turku, Finland \label{in:UTU}
\and
Space Research Institute, Russian Academy of Sciences, Profsoyuznaya 84/32, 117997 Moscow, Russia \label{in:RAS}
\and
Anton Pannekoek Institute for Astronomy, University of Amsterdam, Science Park 904, 1098XH Amsterdam, the Netherlands \label{in:UA}
\and
Department of Physics, P.O. Box 64, 00014 University of Helsinki, Finland \label{in:UH}
\and
Dipartimento di Fisica, Universit\`{a} degli Studi di Roma ``Tor Vergata'', Via della Ricerca Scientifica 1, 00133 Roma, Italy \label{in:UniRoma2} 
\and 
Dipartimento di Fisica, Universit\`{a} degli Studi di Roma ``La Sapienza'', Piazzale Aldo Moro 5, 00185 Roma, Italy \label{in:LaSapienza}
\and
INAF Osservatorio Astronomico di Brera, Via E. Bianchi 46, 23807 Merate (LC), Italia \label{in:OAB}
\and
INAF/IASF Palermo, Via Ugo La Malfa 153, 90146 Palermo, Italy \label{in:IASF}
\and
Dipartimento di Fisica, Universit\`a degli Studi di Cagliari, SP Monserrato-Sestu, KM 0.7, Monserrato 09042, Italy \label{in:UCa}
\and
Institute of Space Sciences (ICE, CSIC), Campus UAB, Carrer de Can Magrans s/n, 08193 Barcelona, Spain \label{in:CSIC}
\and
Institut d’Estudis Espacials de Catalunya (IEEC), 08860 Castelldefels (Barcelona), Spain \label{in:IEEC}
\and
Dipartimento di Fisica e Chimica – Emilio Segrè, Università di Palermo, Via Archirafi 36, 90123 Palermo, Italy \label{in:UPa}
\and 
Mullard Space Science Laboratory, University College London, Holmbury St Mary, Dorking, Surrey RH5 6NT, UK \label{in:UCL}
\and
ASI - Agenzia Spaziale Italiana, Via del Politecnico snc, 00133 Rome (RM), Italy \label{in:ASI}
\and
MIT Kavli Institute for Astrophysics and Space Research, Massachusetts Institute of Technology, Cambridge, MA 02139, USA \label{in:MIT}
\and
INAF Osservatorio Astronomico di Cagliari, Via della Scienza 5, 09047 Selargius (CA), Italy \label{in:OAC}
\and
Astrophysics Science Division, NASA Goddard Space Flight Center, Greenbelt, MD 20771, USA \label{in:Godd1}
\and
Joint Space-Science Institute, NASA Goddard Space Flight Center, Greenbelt, MD 20771, USA \label{in:Godd2}
        }
   \date{Received August 2, 2024; accepted December 7, 2024}
 
\abstract{
We report on the discovery of polarized X-ray emission from an accreting millisecond pulsar. 
During a 10 day-long coverage of the February 2024 outburst of {\srga}, the Imaging X-ray Polarimetry Explorer (\ixpe) detected an average polarization degree of the 2--8~keV emission of $2.3\% \pm 0.4\%$ at an angle of  $59\degr\pm6\degr$ (East of North; uncertainties quoted at the 1$\sigma$ confidence level). 
The polarized signal shows a significant energy dependence with a degree of $4.0\%\pm0.5\%$ between 3 and 6~keV and  $<1.5\%$  (90\% c.l.) in the 2--3 keV range.
We used \nicer, \xmm, and \nustar observations to obtain an accurate pulse timing solution and perform a phase-resolved polarimetric analysis of {\ixpe} data. 
We did not detect any significant variability of the Stokes parameters $Q$ and $U$ with the spin and the orbital phases. 
We used the relativistic rotating vector model to show that a moderately fan-beam emission from two point-like spots at a small magnetic obliquity ($\simeq 10\degr$) is compatible with the observed pulse profile and polarization properties. 
{\ixpe} also detected 52 type-I X-ray bursts, with a recurrence time $\Delta t_{\rm rec}$ increasing from 2 to 8~h as a function of the observed count rate $C$ as $\Delta t_{\rm rec}\propto C^{-0.8}$. 
We stacked the emission observed during all the bursts and obtained an upper limit on the polarization degree of $8.5$\% (90\% c.l.).
}

   \maketitle
%

\section{Introduction} \label{sec:intro}

The simultaneous measurement of the mass $M$ and the equatorial radius $R_{\rm eq}$ of a few neutron stars (NSs) is of paramount importance for constraining the equation of state of ultra-dense matter because of its one-to-one mapping to the NS mass–radius dependence \citep{Lattimer16,ozel16,Baym18}.
Relativistic effects bend the light trajectory and modify energy of the photons emitted by hot spots on the surface of millisecond pulsars  \citep[MSPs; see, e.g., ][and references therein]{watts16}. As a result, modeling the X-ray pulse profiles of MSPs is one of the most powerful techniques to obtain the desired information on $M$ and $R_{\rm eq}$ \citep{poutanen03,miller16}.  The pulse shape also depends on several geometrical parameters (e.g., the binary inclination $i$, the spot co-latitude $\theta$, and its angular size  $\rho$) as well as spectral parameters (e.g., the angular pattern of the emitted radiation; see, e.g., \citealt{poutanen06} and Section~\label{disc:phase}). Breaking this degeneracy with X-ray spectral-timing data requires a very large number of counts \citep{lo13,miller15}. A few megasecond-long observations of rotation-powered MSPs with the Neutron Star Interior Composition Explorer Mission (NICER) met this requirement and led to constraints on $M$ and $R_{\rm eq}$ with a relative uncertainty of $\sim 10\%$ for three of them \citep{riley19,riley21,miller19,miller21,choudhury24,dittmann24,salmi24,vinciguerra24}.

Accreting millisecond pulsars \citep[AMSPs;][]{patruno21,disalvo22} represent an intriguing alternative.  During accretion outbursts, they attain an X-ray luminosity of $L_{\rm X}\simeq 10^{36}$--$10^{37}$~\lumcgs, i.e. 5--6 orders of magnitude brighter than rotation-powered MSPs. 
The degeneracy between $M$ and $R_{\rm eq}$ and the spot geometrical and spectral parameters \citep[see, e.g.,][]{viironen04} prevented to get meaningful constraints from the data taken with, e.g., the  Rossi X-ray Timing Explorer   ($M=$1.2--1.6 ${M_\odot}$, $R_{\rm eq}=$6--13~km by \citealt{poutanen03}; $M=$0.8--1.7${M_\odot}$, $R_{\rm eq}=$5--13~km by \citealt{morsink11}). 
However, \citet{salmi18} showed that if $i$ and $\theta$ were known a priori with an accuracy of a few degrees, those data would have allowed measurements with a relative uncertainty 
comparable to or slightly smaller than that for rotation-powered MSPs.  

X-ray polarimetry might provide key information on the geometrical parameters of the spots. Soft photons from the NS surface are Compton up-scattered by hot electrons in the accretion shock \citep{poutanen03} achieving polarization degree (PD) from a few per cent up to 10--20\%  \citep{nagirner94, viironen04,salmi21,bobrikova23}. 
For the magnetic field strength of AMSPs ($10^8$--$10^9$~G), the typical energy of X-ray photons is orders of magnitude larger than the fundamental cyclotron energy ($\simeq 1$--$10$~eV), and scattering is not affected by the magnetic field. 
As a result  of azimuthal symmetry, the polarization vector of the X-rays emitted from the accretion shock is expected to lie either in the meridional plane, formed by the photon momentum and the normal to the surface, or perpendicular to it \citep{viironen04,poutanen20}. 
As the NS rotates, the polarization angle (PA) should thus swing as a function of the pulsar spin phase in accordance with the rotating vector model \citep[RVM;][]{rvm69,Meszaros88}. 
In  AMSPs, the relativistic motion of the emission regions causes additional rotation of the polarization plane \citep{poutanen20}; also the oblate shape of a quickly rotating NS has to be accounted for \citep{loktev20}.

\begin{figure}
\includegraphics[width=8.5cm]{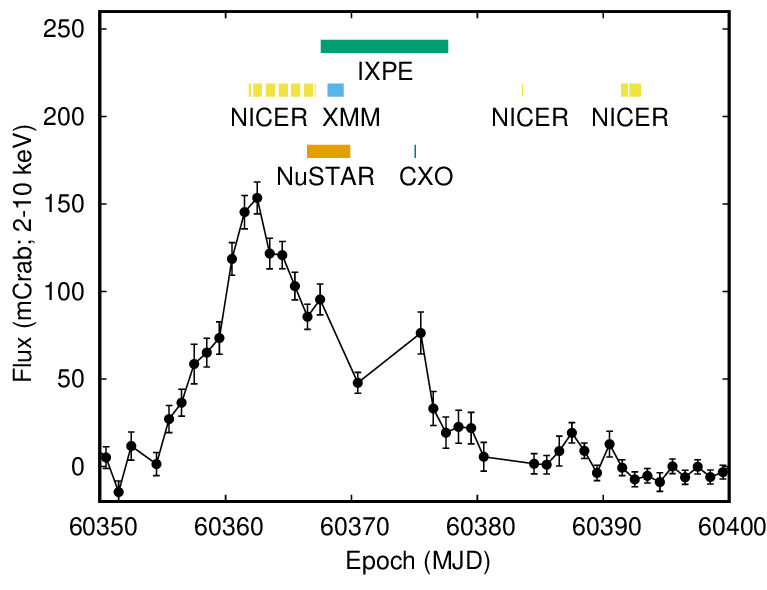}
\caption{Light curve of the 2024 outburst of SRGA~J1444 observed by MAXI \citep{Matsuoka2009}. We converted 2--20~keV observed count rates into 2--10~keV flux values assuming that the spectrum is described by a power law with a photon index $\Gamma=1.9$ absorbed by an equivalent hydrogen column of $N_{\rm H}=2.9\times10^{22}$~cm$^{-2}$ \citep{ng24}. Horizontal bars indicate the time intervals covered by observations of the instruments discussed in this paper.  \label{fig:lc}}
\end{figure}

Here, we report on the discovery of polarized X-ray emission from the AMSP {\srga} (SRGA~J1444 in the following)  by the Imaging X-ray Polarimetry Explorer (\ixpe; \citealt{ixpe22}) in the context of a campaign which included also {\nicer}, X-ray Multi-Mirror Mission ({\xmm}), and Nuclear Spectroscopic Telescope Array ({\nustar}) observations. Discovered in outburst on 2024 February 21 by the {\it Mikhail Pavlinsky} ART-XC telescope on-board the {Spectrum-Roentgen-Gamma} observatory \citep[SRG, ][]{molkov24} at a relatively bright 4--12~keV X-ray flux of $2\times10^{-9}$~{\fluxcgs} (see Fig.~\ref{fig:lc}), SRGA~J1444 showed coherent pulsations at 447.9~Hz first detected by {\nicer}, coming from a NS in a 5.2~h orbit around a donor with a mass in the range 0.2--0.7${M_\odot}$ \citep{ng24}. SRGA~J1444 also showed type-I X-ray bursts \citep{mariani24} quite regularly, with a recurrence time increasing from 1.6 to 2.2~h  as the X-ray flux was observed to decrease \citep{molkov24}. No optical counterpart could be identified \citep{baglio24} at the most accurate X-ray position obtained with High Resolution Camera observations on board the {\chandra} X-ray observatory \citep[CXO, ][]{illiano24}, most likely due to the extinction of $A_{\rm V}\simeq 7$~mag \citep[estimated\footnote{see the tool available at \url{http://astro.uni-tuebingen.de/nh3d/nhtool}} using the maps of][at a distance of 8~kpc]{doroshenko24}  in that direction. \citet{russell24} reported a radio counterpart with a flux density of $\sim 200$~$\mu$Jy at 5.5~GHz and a flat spectral shape, consistent with a compact jet or discrete ejecta.

This paper is organized as follows. Section~\ref{sec:obs} details the analysis of the different X-ray datasets considered. Section~\ref{sec:persistent} presents the properties of the X-ray persistent\footnote{Here, persistent refers to the accretion-powered X-ray emission seen during the outburst, as opposed to the burst emission.}  emission. The measurement of polarized emission is presented in Sect.~\ref{sec:polanalysis} and its interpretation is discussed in Sect.~\ref{disc:avg}. Section~\ref{sec:timing} reports the pulsar timing solution obtained from the analysis of the coherent X-ray pulsations observed by the various instruments involved in this analysis, Sect.~\ref{sec:phase-resolved} the observed phase-resolved variability of the polarized emission, and Sect.~\ref{disc:phase} its modeling.  Section~\ref{sec:bursts}  outlines the properties of recurrence  of the type-I X-ray bursts detected by IXPE, Sect.~\ref{sec:burstpol} its the X-ray polarization and Sect.~\ref{disc:burst} the physical implications. The main results of this paper are summarized in Sect.~\ref{conclusions}.

\section{Observations}
\label{sec:obs}

\begin{table*}
\centering
\caption{Observations analyzed in this paper.
\label{tab:log}}
\begin{tabular}{lcccccc}
\hline\hline
Observatory & Obs. ID & Start date (MJD) & End Date (MJD) & Instrument & Exposure (s) & Energy band (keV)\\
            \hline
\ixpe & 03250101  & 60367.556 & 60377.676 & DU1 & 552930 & 2--8 \\
    &   &   &   & DU2 & 553541 & \\
      &   &   &   & DU3 & 554066 & \\
\nicer & 620419 &  60361.831 & 60363.908 & XTI & 9536 & 0.3--10 \\
 & 663908 &  60364.219 & 60392.971 & & 24805 & \\ 
 \nustar & 80901307002 &  	60366.459 & 60369.883 & FPMA & 157675 & 3--79 \\
  &  &  	 &  & FPMB & 157440 & \\
 \xmm & 0923171501 & 60368.102 & 60369.376 & EPIC-pn & 110077 & 0.3--10 \\ \hline
\end{tabular}
\end{table*}

\begin{figure}
\includegraphics[width=8.5cm]{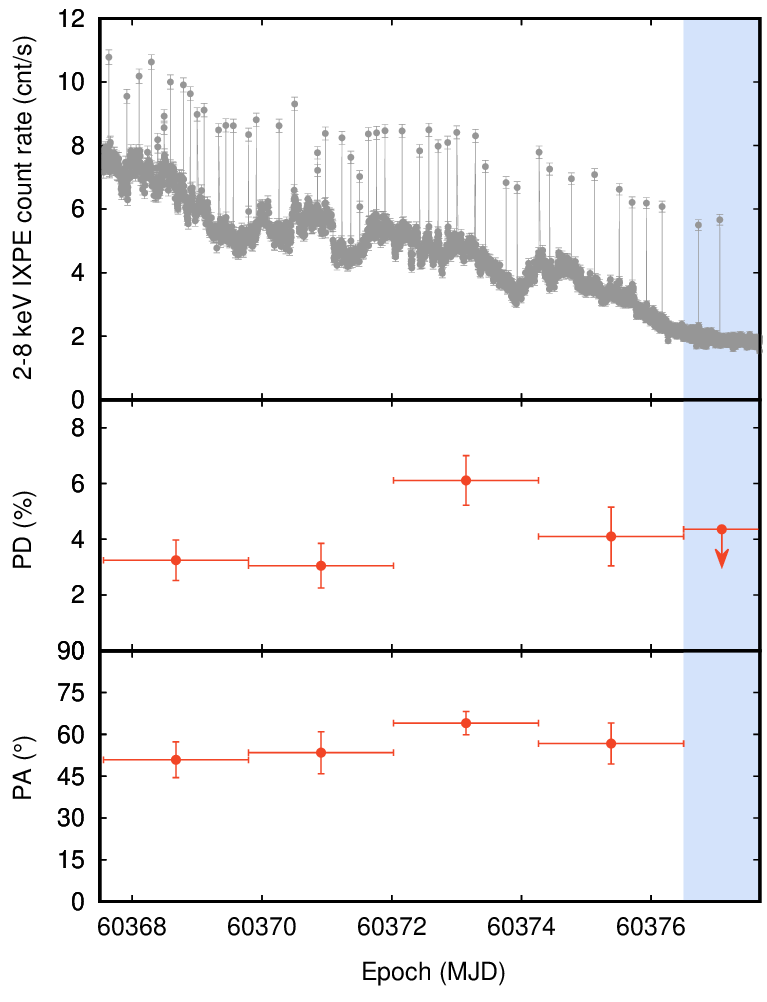}
\caption{IXPE 2--8 keV light curve in 200~s time bins (top panel). Type-I X-ray bursts are easily recognized. PD (middle panel) and PA (bottom panel)  of the total (i.e., persistent and bursting) emission measured in the  3--6~keV band and evaluated over $\sim 2$~d-long intervals. The blue shaded region indicates the interval after the pulse amplitude increase observed at MJD 60376.5 in which a 90\% c.l. upper limit on the PD is reported.
\label{fig:lc_pol}}
\end{figure}

Table~\ref{tab:log} lists the observations analyzed in this paper, using \textsc{HEASoft} package version 6.33.2. {\ixpe} started observing {SRGA~J1444} less than a week after the first detection of the source, in response to the trigger of the General Observer (GO) program 03250101 (PI: A.~Papitto). 
Observations lasted $\sim$10~d with visibility gaps giving a duty cycle exceeding 60\%.
We used the \texttt{pcube} algorithm in the \texttt{xpbin} tool of the \textsc{ixpeobssim} package version 30.6.4 \citep{Baldini2022}, along with the {\ixpe} calibration database (CALDB) released on 2024 February 28, to extract polarimetric information using the formalism from \citet{kislat15}. We selected source photons from a circular region with a radius of 100\arcsec\ centered at the source position. We use the unweighted analysis implemented in \textsc{ixpeobssim} \citep{DiMarco_2022} to measure the Stokes parameters $(I,Q,U)$. Figure~\ref{fig:lc_pol} shows the 2--8~keV IXPE count rate, the polarization degree PD=$\sqrt{q^2+u^2}$ and angle PA=$\frac{1}{2} \arctan (u/q)$ measured in the 3--6~keV band, where a significant polarization is detected (see Sect.~\ref{sec:polanalysis} and Fig.~\ref{fig:enpol}). Here, $q=Q/I$ and $u=U/I$.
As the source count rate is relatively high ($\sim$2--8 count~s$^{-1}$), the background contributes at most 5\% of the total rate and we did not subtract it following the recommendation of \citet{DiMarco_2023}. For the {\ixpe} spectral analysis, we used the response matrices \textsc{20240101\_alpha075} present in the CALDB, computed ancillary response files using \texttt{ixpecalcarf} weighting the Stokes parameters following the analysis presented by \citet{DiMarco_2022} and binned the $I$ spectra to have at least 30 counts per bin. To ensure statistics high enough for a spectro-polarimetric analysis, we binned data to a constant resolution of 200 eV.

\citet{ng24} reported the analysis of NICER observations during the first three days of the outburst using publicly available data. Here, we also include data taken afterwards in the GO program ID 663908 (PI: Papitto). Visibility constraints and high background prevented {\nicer} to collect data simultaneously with the {\ixpe} (see Fig.~\ref{fig:lc}). The monitoring briefly resumed on 2024 March 14 and more regularly after March 21, when the source count rate had already decreased close to the background value. We reduced the observations using calibration version \textsc{xti20240206} and the same filtering criteria adopted by \citet{ng24}, retaining photons with energy ranging between 0.3 and 10~keV for a pulse timing analysis.

The European Photon Imaging Camera (EPIC) pn-CCD camera on-board {\xmm} observed SRGA J1444 uninterruptedly for $\sim$110~ks.
We retained 0.3--10~keV source photons extracted from a 21 pixel-wide stripe (equivalent to 86\farcs1) around the position of the pulsar to perform a pulse timing analysis.

{\nustar} provided the longest overlap with {\ixpe} data starting on 2024 February 26.
We reduced the data using the calibration version \textsc{indx20240325}, with the fine clock correction file ver.~192. A circular region with a wide radius of 140\arcsec\ was adopted to extract source photons in the 3--79~keV band.

The times of arrival of the photons recorded by all the instruments were transformed into the inertial reference frame centred at the Solar System barycenter using the Chandra coordinates reported by \citet{illiano24}, R.A.= 14$^{\rm h}$ 44$^{\rm m}$ 58\fs94, Dec = $-$60\degr 41\arcmin 55\farcs3 (J2000) and the DE 405 ephemeris.  We detected 52, 5, 13 and 23 type-I X-ray bursts during {\ixpe}, {\nicer}, {\xmm} and {\nustar} exposures, respectively. When analysing the {\it persistent} emission, we discarded intervals between 5-s before and 60-s after the onset of each burst.

\section{Persistent X-ray emission}
\label{sec:persistent}

\subsection{Phase-averaged emission}

\subsubsection{Average polarization}
\label{sec:polanalysis}

\begin{figure}
\includegraphics[width=8.5cm]{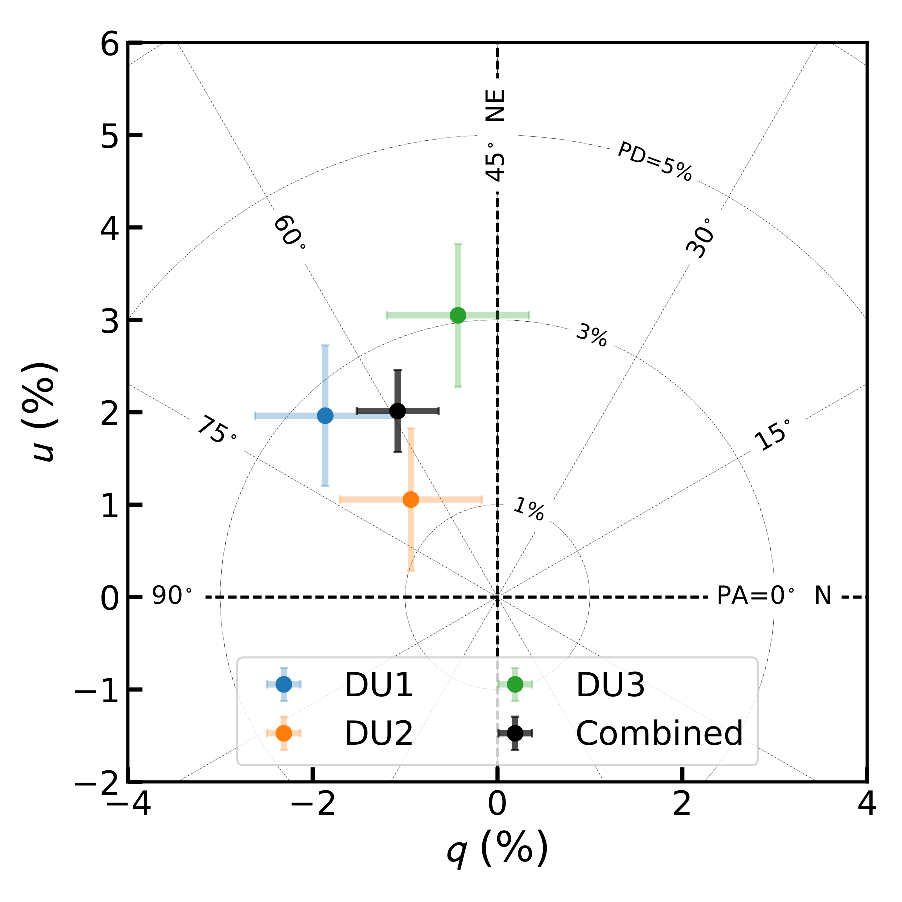}
\caption{Average normalized Stokes parameters observed during the 2--8~keV {\it persistent} emission for the single IXPE detector units (DUs) and for the sum. 
\label{fig:avgpol}}
\end{figure}

\begin{figure}
\includegraphics[width=8.5cm]{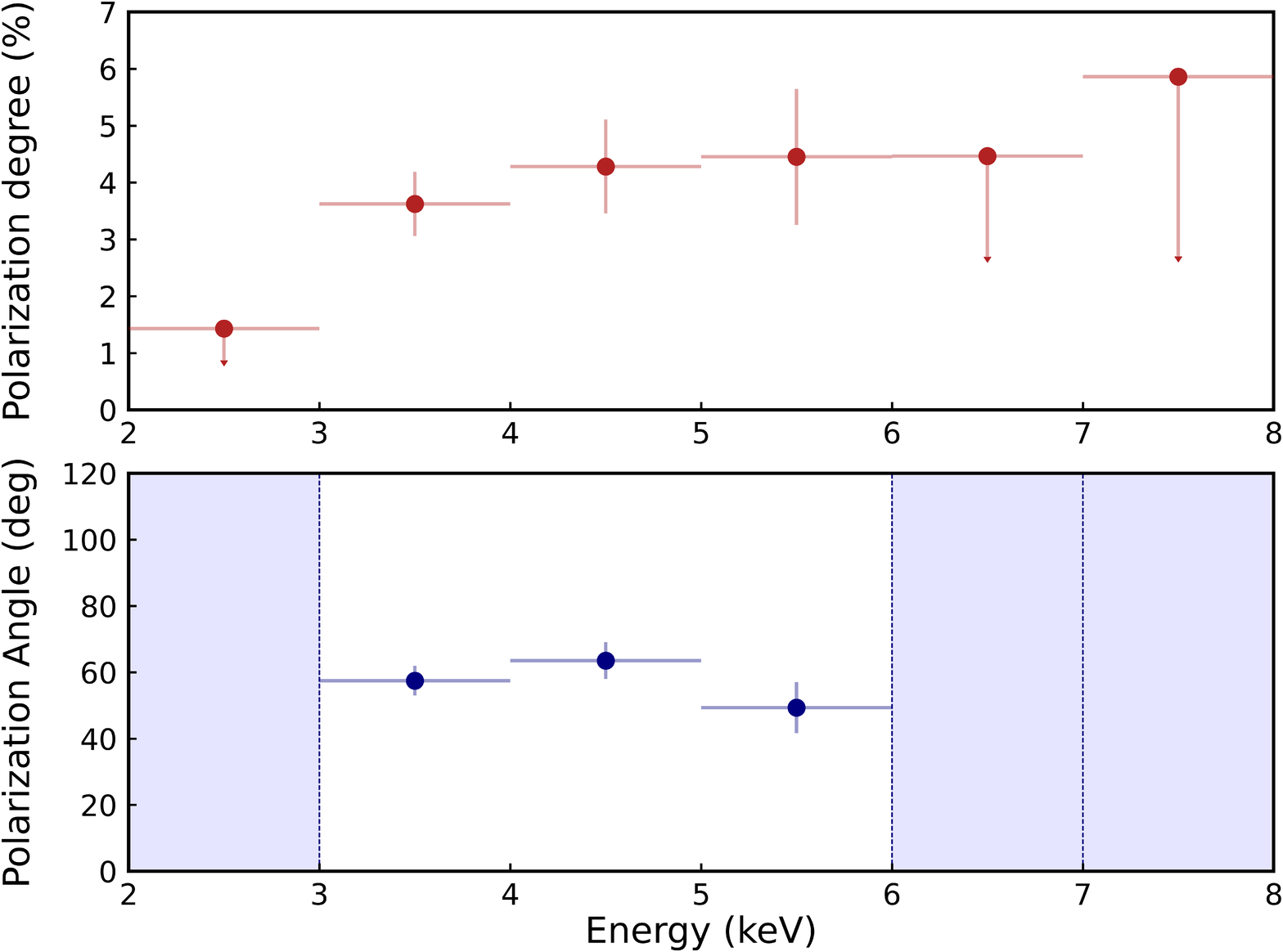}
\caption{PD (top) and PA (bottom) as a function of energy. 
Above 6 keV,  polarization is compatible with zero and we report upper limits at 90\%~c.l., while the PA is unconstrained. \label{fig:enpol}}
\end{figure}

Figure~\ref{fig:avgpol} shows the average normalized Stokes parameters observed during the {\it persistent} emission by the three IXPE detector units. The average Stokes parameters are $q=-1.1\%\pm0.4\%$ and $u=2.0\%\pm0.4\%$, which translate into an average PD of  $2.3\% \pm 0.4$\%, and an angle of PA $59\degr\pm6\degr$  (measured East of North, errors are quoted at the 1 $\sigma$ c.l. unless differently specified; see Fig.~\ref{fig:avgpol}). The probability of detecting such values of the normalized Stokes parameters from noise fluctuations is $1.1\times10^{-7}$ (corresponding to a 5.3 $\sigma$ c.l.).
Figure~\ref{fig:enpol} shows the results of an energy-resolved analysis. Polarized emission is detected at the highest statistical significance in the 3--6~keV energy range with an average  PD$_{3-6}=4.0\%\pm0.5\%$ and PA$_{3-6}=57\degr\pm3\degr$. Between 2 and 3~keV, the PD is not significant ($0.7\%\pm0.6\%$) with an upper limit at 90\% confidence level of PD$_{2-3}<1.5\%$.
Above 6~keV, the 5\% upper limit on the PD
becomes larger than both PD$_{3-6}$ and the minimum detectable polarization at 99\% confidence level in the 3--6~keV band (MDP$_{99}$=1.4\%). 
We did not detect any significant variation or rotation of the 3--6~keV polarization vector over 2~d-long intervals, even though the PD slightly increases halfway through the observation  (see bottom panels of Fig.~\ref{fig:lc_pol}). We checked that no significant variation of the spectral parameters occurs in the time interval characterized by a slightly larger PD compared to the emission observed elsewhere.

Similar to \citet{ng24}, we used the sum of a black-body (\texttt{bbodyrad} in \textsc{xspec}) and thermal Comptonization  (\texttt{nthcomp}) to satisfactorily model ($\chi^2=483.9$ for $441$ d.o.f.) the 2--8 keV {\ixpe} energy spectrum. We modelled absorption with the Tuebingen-Boulder model \citep{wilms00} with  the absorption column fixed to $N_{\rm H}=2.9\times10^{22}$~cm$^{-2}$
and quote uncertainties on the spectral parameters at 90\% c.l.
The thermal component has a temperature of $kT_{\rm bb}=0.52\pm0.01$~keV and an apparent size of $R_{\rm bb}=(14.0\pm0.5) d_{8}$~km, where $d_{8}$ is the distance to the source in units of 8~kpc. Comptonization is described by an asymptotic power law with index $\Gamma=1.70\pm0.03$ and a soft photon input with temperature $<0.15$~keV. The electron temperature was fixed at a value beyond the {\ixpe} band-pass $kT_{\rm e}=15$~keV. The Comptonization component accounts for $\sim95\%$ of the 2--8 keV observed flux of ($5.6\pm0.1)\times10^{-10}$ \fluxcgs.
Spectral analysis of the {\nicer} observation (Obs.~ID 6639080103) performed $\sim$12~h before the start of the {\ixpe} coverage indicates a smaller black-body size  ($kT_{\rm bb}=0.68\pm0.05$~keV; $R_{\rm bb}=(5.7\pm1.0) d_{8}$~km) and a  steeper slope of the Comptonized emission ($\Gamma=1.90\pm0.01$). 

The decrease of the PD below 3 keV, i.e. where the soft thermal component gives its highest contribution to the 2--8~keV spectrum, motivated us to perform a spectro-polarimetric analysis. 
By fitting the spectra of the three Stokes parameters with a 
constant polarization component to multiply the model presented above,  \texttt{tbabs*(bbodyrad+nthcomp)*polconst}, gave  PD=$2.6\%\pm0.5\%$ and PA=$59\degr\pm6\degr$ with a $\chi^2$/d.o.f. of 340/257, compatible with the \texttt{pcube} results.
By using separate constant polarization for the two spectral components gave a slight improvement of the fit $\chi^2$/d.o.f.=328/255   
with a \texttt{bbodyrad} component polarized with the PD=$8\%\pm5\%$ at PA=$-43 \degr \pm 19\degr$ and the \texttt{nthcomp} with the PD=$4.6\%\pm 1.1\%$ and PA=$55\degr \pm 7\degr$.

\subsubsection{Interpretation}
\label{disc:avg}

For the first time, {\ixpe} observations of \mbox{SRGA J1444} allowed us to significantly detect polarization from an AMSP with an average 2--8~keV degree of $2.3\%\pm0.4\%$. Such a value increases to $4.0\%\pm0.5\%$ if the analysis is restricted to the 3--6~keV energy band.   Comptonization by hot ($kT_{\rm e} \sim$20--60~keV) electrons in slabs of moderate optical depth ($\tau\approx$1--2) located above hot spots ($kT_{\rm bb}\sim 1$~keV)  on the NS surface accounts for more than 90\% of the emission in the {\ixpe} bandpass of AMSPs such as SRGA J1444 \citep{poutanen06b}  and represents the most natural explanation of the polarization we detected.  Values of the PD as high as 10--20\% were predicted by modelling the scattering in Thomson approximation \citep{viironen04,salmi21}. However, the actual degree of polarization is a function of the physical properties of the scattering medium \citep{nagirner94}.
Recently, \citet{bobrikova23} applied the formalism for Compton scattering in a hot slab \citep{nagirner93,poutanensvensson96} to describe the Stokes parameters as a function of energy and angle in terms of the electron temperature and optical depth of the slab and the seed photon temperature. They found values of the PD of a few per cent at the most, decreasing to values undetectable to {\ixpe} at high ($>50$~keV) electron temperatures. They concluded that chances for {\ixpe} to detect polarization from an AMSP significantly increased for systems with a larger optical depth ($\tau>2$) and lower electron temperature ($kT_{\rm e} < 30$~keV).  \citet{molkov24} fitted the 4--30~keV ART-XC spectrum with Comptonization from a medium with $kT_{\rm e}\sim 25$~keV and $\tau\sim 3.2$ (evaluated from the photon index of $\Gamma=1.8$ as in, e.g., \citealt{lightman87}). Preliminary fitting of the 3--79~keV {\nustar} spectrum (Malacaria et al., in prep.) suggests an even lower temperature ($kT_{\rm e}=10.9^{+0.2}_{-0.1}$~keV) and higher optical depth ($\Gamma=2.2\pm0.1$, $\tau=4.0\pm 0.3$).  In the framework developed by \citet{bobrikova23}, SRGA J1444 may thus present favorable spectral conditions to produce a relatively high degree of X-ray polarization compared to other AMSPs which might show even lower values. In addition, \citet{bobrikova23} showed that the observed polarization degree increased when an observer mostly sees the spot under high angles (i.e., $|i-\theta|$ close to $90\degr$). Such a configuration is supported by the results presented in Sect.~\ref{sec:phase-resolved}.

We detected a significant decrease in the PD  below 3 keV. In the IXPE bandpass, that energy band is where the non-upscattered blackbody emission contributes the most to the total observed flux. These photons are not expected to be polarized and might explain the decrease of the total PD. However, the soft thermal component accounts for just 5--10\% of the total emission ($7.7\%\pm1.4\%$ between 2 and 3~keV, according to the {\nicer} spectral analysis reported in Sect.~\ref{sec:polanalysis}). As a result, the influence of such an unpolarized component at low energies does not readily explain the observed decrease of the PD. The analysis presented in Sect.~\ref{sec:polanalysis} shows that, to explain the decrease of the total polarization, the soft thermal component should be polarized to a degree of $8\% \pm 5\%$ and its polarization angle should differ by $\sim 90\degr$ from the polarization angle compared to the Comptonization component. However, the improvement in the fit description compared to a model with the same polarization properties for the two spectral components is modest. As thermal unscattered emission is not expected to be polarized, an intrinsic decrease of the polarization degree of the Compton scattered photons is a more likely explanation of the observed energy dependence. \citet{bobrikova23} predicted a swing by $\sim$90\degr\  of the polarization vector of the Comptonized emission at an energy of $\simeq$5~keV,  only slightly higher than the turn-over energy of $\simeq$3~keV we observed. A comprehensive analysis of the spectral properties of the source, including also {\xmm} data, is in preparation.

\subsection{X-ray pulsed variability}

\subsubsection{Pulse phase timing solution}
\label{sec:timing}

\begin{figure}
\includegraphics[width=8.5cm]{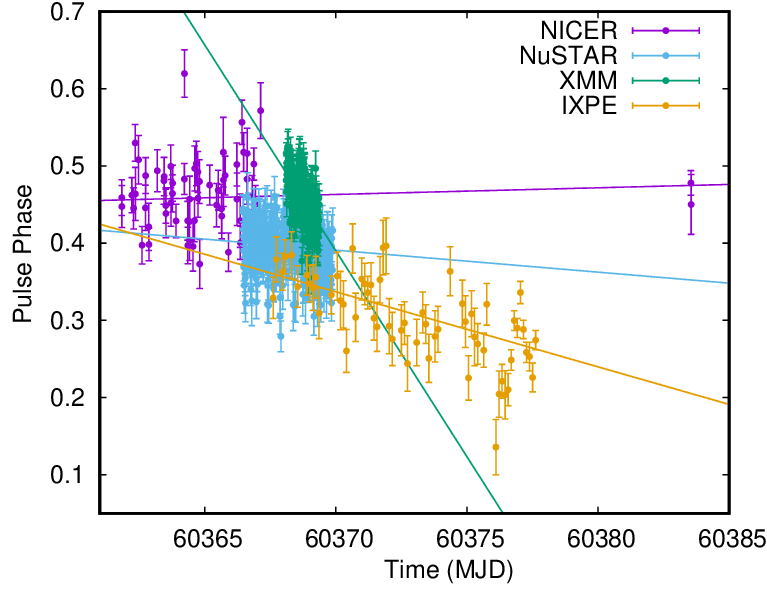}
\caption{Pulse phase computed over the first harmonic of the pulses observed by {\nicer} (magenta), {\nustar} (light blue), {\xmm} (green) and {\ixpe} (yellow) by folding 500~s-long intervals (5~ks for {\ixpe}) around $\nu_{\rm fold}$ (see Table~\ref{tab:timing}) with a reference epoch $60361.8$~MJD. Solid lines mark the best-fit linear trends.
\label{fig:timing}}
\end{figure}

We used the pulsar timing solutions based on the first few days of {\nicer} observations \citep{ng24} to correct the arrival time of X-ray photons for the light travel time delays due to the pulsar orbit.
We used two harmonic components to model the pulse profiles obtained by folding data taken in $\sim$500~s-long intervals and sampled by 16 phase bins. Table~\ref{tab:timing} shows the timing solutions based on the observed evolution of the phase computed over the fundamental harmonic for the various instruments considered here, plotted in Fig.~\ref{fig:timing} with the best-fitting linear trend). We attribute the residual random phase variability by $\sim$0.05--0.1 cycles to pulse timing noise, as observed from other AMSPs with an even larger magnitude \citep{patruno21,disalvo22}. The two values of the pulse phase measured by {\nicer} at MJD 60384,  after an interruption of the monitoring due to visibility constraints, line up with the trend observed earlier and allow us to put a tight upper limit on the spin frequency derivative ($\dot{\nu}<1.2\times10^{-13}$~Hz~s$^{-1}$). However, we caution that such an alignment could be fortuitous, as the pulse phase of AMSPs often shows erratic variability, especially in the late stages of an outburst \citep[see, e.g.][]{illiano23}.

\begin{table*}
\centering
\caption{Pulse phase timing solutions of \mbox{SRGA J1444}.
\label{tab:timing}}
\begin{tabular}{lcccc}
\hline\hline
 & \nicer & \xmm & \nustar & \ixpe\\ \hline
Interval (MJD) & 60361.837--60383.544 & 60368.106--60369.374 & 60366.465--60369.881 & 60367.615--60377.619 \\
$\phi(T_0)$ & $0.47(1)$ & $0.500(7)$ & $0.400(5)$ & $0.37(1)$\\
$\nu(T_0)$ (Hz) & $447.87156112(2)$ & $447.8715620(3)$ & $447.87156127(2)$ & $447.87156138(8)$  \\
$\dot{\nu}$ (Hz\,s$^{-1}$) & $<1.2\times10^{-13}$ & $<7.0\times10^{-12}$ & $<1.5\times10^{-12}$ & $<3.0\times10^{-13} $ \\
$a_1 \sin i /c$ (lt-s) & $0.65052(2)$ & $0.650491(5)$ & $0.650498(3)$ & 0.650486(13) \\
$P_{\rm orb}$ (s) & $18803.65(1)$ & $18803.65(1)$ & $18803.673(4)$ &  $18803.665(4)$ \\
$T_{\rm asc}$ (MJD) & $60361.641306(3)$ & $60361.641305(6)$ & $60361.641297(2)$ & 60361.641310(5) \\
$e$ & $<9\times10^{-5}$ & $<9\times10^{-5}$ & $<3\times10^{-5}$ & $<1.0\times 10^{-4}$ \\
$\chi^2$ (d.o.f.) &  93.0 (53) & 368.5 (212) & 1261.8 (762) & 121.3 (53) \\ \hline
\end{tabular}
\tablefoot{All the solutions presented in the table were obtained by folding data around $\nu_{\rm fold}=447.87156116029$~Hz with a reference epoch MJD 60361.8. To facilitate the comparison, the frequency estimates were evaluated at $T_0=$MJD~60368.0. All the epochs reported here are in the Barycentric Dynamical Time (TDB) system. Uncertainties at the 1$\sigma$ confidence level are reported in parentheses. Upper limits are given at a 90\% confidence level.}
\end{table*}

\begin{figure}
\includegraphics[width=8.5cm]{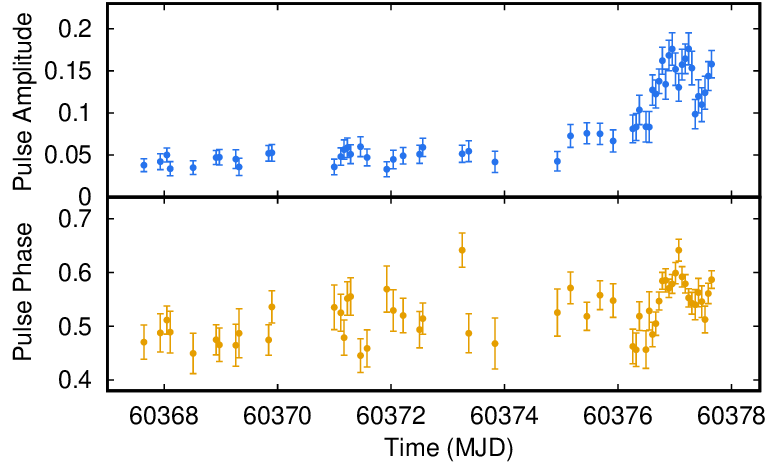}
\caption{Pulse fractional amplitude and phase of the first harmonic component used to model the pulse profile obtained by folding {\ixpe} data into 5~ks-long interval around the timing solution reported in the rightmost column of Table~\ref{tab:timing}.}
\label{fig:ixpetiming}
\end{figure}

We used these ephemerides to fold 5~ks-long intervals of {\ixpe} data and recover the pulsed signal despite the lower counting statistics. Figure~\ref{fig:ixpetiming} shows the fractional amplitude and phase of the fundamental harmonic observed by {\ixpe}. An increase of the amplitude from $\sim4$ to 15\%  is observed after MJD 60376.5.

The frequencies measured by {\nustar}, {\ixpe} and {\xmm} 
differ by $\delta\nu/\nu\simeq 0.3\times10^{-9}$, $0.6\times10^{-9}$ and $2\times10^{-9}$ with respect to the NICER measurement, respectively. The latter is compatible with the value quoted in the latest calibration technical note for the {\xmm} clock relative time accuracy\footnote{\url{https://xmmweb.esac.esa.int/docs/documents/CAL-TN-0018.pdf}} ($\delta\nu/\nu\simeq 10^{-8}$).
The pulse phase recorded by {\nustar} and {\ixpe} at epoch $T_0$ precedes those observed by {\nicer} by 0.07--0.1 cycles. The NuSTAR shift can be understood in terms of the dependence of the pulse phase on energy already noticed by SRG/ART-XC \citep{molkov24}; the reason for the {\ixpe} mismatch is less clear and will be addressed in a future paper. For consistency, the phase-resolved analysis presented in Sect.~\ref{sec:phase-resolved} is based on the {\ixpe} timing solution, even if slightly less accurate.

\subsubsection{Phase-resolved polarization variability}
\label{sec:phase-resolved}

To perform a phase-resolved analysis of the polarimetric properties of SRGA J1444, we considered {\ixpe} data corrected with the orbital parameters listed in Table~\ref{tab:timing}.  We used the \texttt{pcube} algorithm to evaluate in $n=16$ phase bins the Stokes parameters $Q$ and $U$, which are expected to be normally distributed \citep{gonzalez23,suleimanov23}. We restricted the analysis to the 3--6~keV energy band in order to maximize the strength of the polarized signal.  
We used the \texttt{pcube} algorithm without acceptance corrections to measure the Stokes parameters in counts per second.

\begin{figure}
\includegraphics[width=8.5cm]{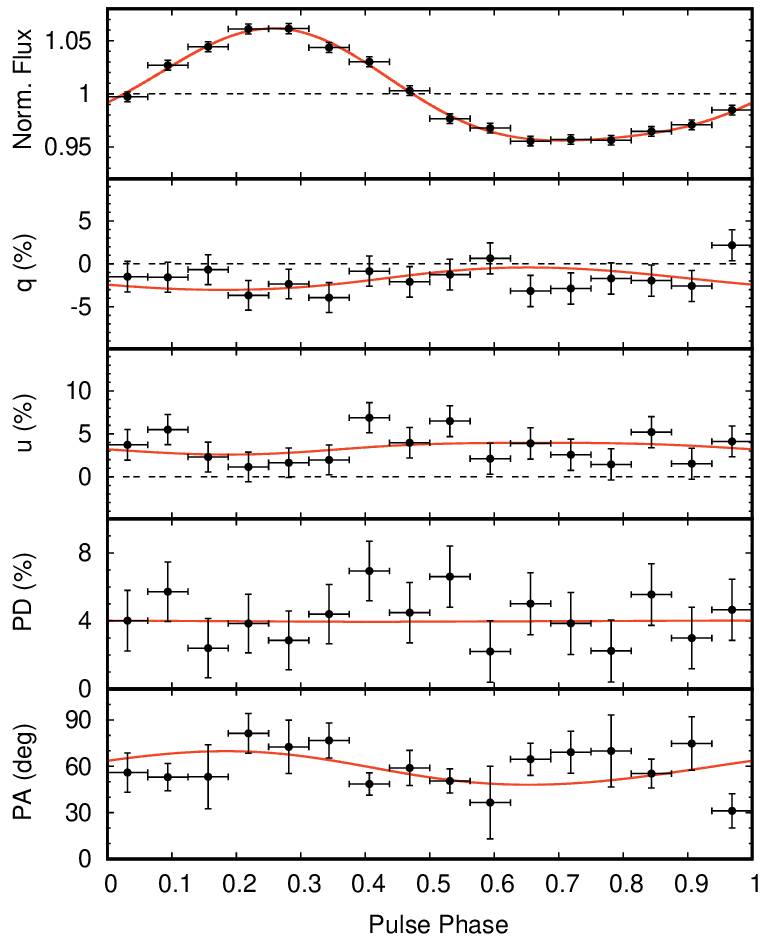}
\caption{Pulse phase dependence of polarimetric properties. From top to bottom: total flux normalized to the average, normalized Stokes parameters $q=Q/I$ and $u=U/I$, the PD and PA observed by IXPE during the interval MJD 60367.6--60376.5, as a function of the spin phase. The solid red line indicates the model obtained {\bf by} fitting the absolute Stokes parameters assuming two point-like spots and the parameters listed in text.  
\label{fig:stokes_intA}}
\end{figure}

Figure~\ref{fig:stokes_intA} shows the phase-resolved  Stokes parameters, the PD, and the PA observed before the pulse amplitude increase which occurred at MJD 60376.5.
The Stokes parameters $Q$ and $U$ did not show significant variability. A fit with a constant function gave a $\chi^2$ of 16.6 and 15.6 for 15 d.o.f., respectively, with  average values of $q=Q/I=-1.6\%\pm0.5\%$ and  $u=U/I=3.5\%\pm0.4\%$.

\begin{figure}
\includegraphics[width=8.5cm]{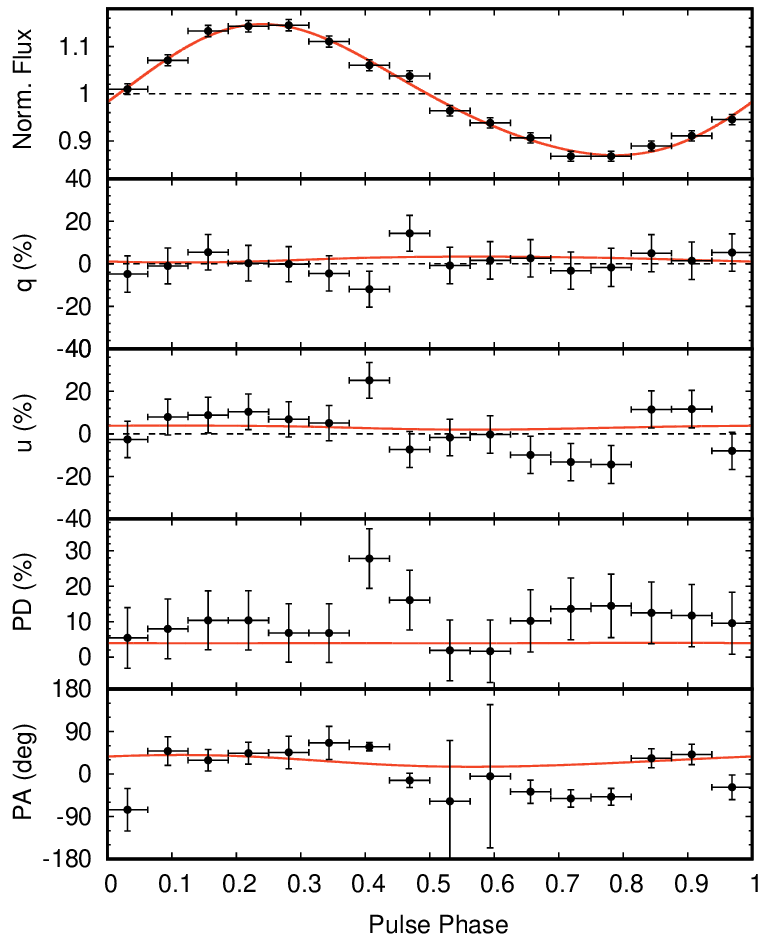}
\caption{Same as Fig.~\ref{fig:stokes_intA} but for the interval MJD 60376.5--60377.7 only. \label{fig:stokes_intB}}
\end{figure}

We also analyzed separately data taken after the pulse amplitude increase which occurred at MJD 60376.5 (see Fig.~\ref{fig:stokes_intB}). 
Fitting with a constant gives $q'=0.5\%\pm1.4\%$ and $u'=2.4\%\pm2.7\%$ for $\chi^2=10.2$ and 18.7 for 15 d.o.f. Both Stokes parameters are compatible with zero within the uncertainties, even though the value of $u'$ measured at phase 0.4 significantly deviated from the trend, leading to an increase of the PD up to $\sim30$\%.

We also measured the Stokes parameters over 15 orbital phase bins to search for orbital variability of the polarization properties but found none significant. Constant values of $Q$ and $U$ modeled well the dependence of these values on the orbital phase ($\chi^2=9.1$ and 12.5 for 14 d.o.f., respectively). 

\subsubsection{Modelling of the phase-resolved polarization properties}
\label{disc:phase}

\begin{figure}
\includegraphics[width=8.5cm]{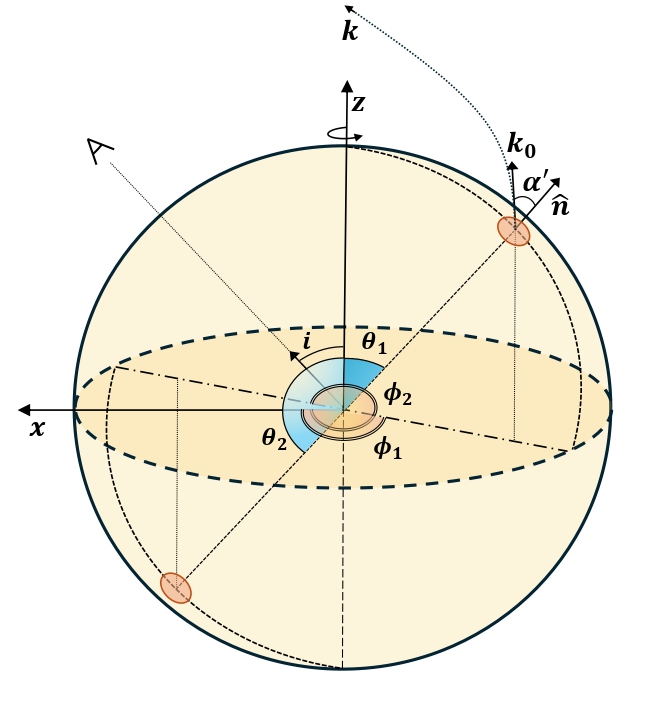}
\caption{Assumed geometry of the problem, adapted from \citet{poutanen06}. The binary inclination $i$, the co-latitutde of the two spots $\theta_1$ and $\theta_2$, their phase $\phi_1$ and $\phi_2$ and the angle made by the direction of the emitted photon with respect to the normal to the spot in the co-moving frame $\alpha'$ are explictly indicated. \label{fig:geometry}}
\end{figure}

Assuming the accretion columns as the source of the polarized X-ray photons we observed, it is necessary to ask what spot geometrical configuration can explain at the same time the observed pulse profile and the non-significant phase-resolved variability of the Stokes parameters $Q$ and $U$.

Unlike the case of highly magnetized, slowly rotating X-ray pulsars \citep{doroshenko22,tsygankov22,mushtukov23,marshall22,forsblom23,tsygankov23,malacaria23,doroshenko23,suleimanov23,poutanen24,forsblom24}, the polarization properties of AMSPs are defined by the geometry of the spots, relativistic effects \citep{viironen04,poutanen20,loktev20}, and possibly eclipses by the inner parts of the accretion disks \citep{poutanen09,ibragimov09}. This does not allow us to use a simple rotating vector model to fit the observed phase-resolved PD and PA, but requires a fit of the total Stokes vector $(I,Q,U)$ which is the sum of the Stokes vectors of the two spots. 

The spots' Stokes parameters depend on how the total and polarized radiation vary with the spin phase, photon energy and zenith angle \citep{viironen04,poutanen20,salmi21,bobrikova23}. A modeling that includes an accurate description of all these effects is beyond the scope of this letter and will be presented elsewhere.
Here, we draw considerations on the system's geometry by using the approximate formulae given by \citet{poutanen06} and \citet{poutanen20} for the phase-resolved Stokes parameters. We assume two point-like spots on the surface of a rapidly rotating NS (see Fig.~\ref{fig:geometry} for a cartoon of the assumed geometry) which emit X-rays with a power-law spectrum with $I(\alpha',E)=I_0 (1+h\cos{\alpha'}) E^{-(\Gamma-1)}$. Here,  $\Gamma=1.7$ is the spectral photon index, $\alpha'$ is the zenith angle (i.e., the angle between the normal to the spot and emission direction) in the co-rotating frame and $h$ is an anisotropy parameter ($h=0$ for black-body emission, $-1<h<0$ for a fan-beam pattern such that expected from Comptonization in a thin slab; \citealt{poutanen03}). We used Eq.~(20--22) of \citet{poutanen06} to determine the phase intervals in which the photons are visible to the observer and Eqs.~(28--34) of the same paper to express the flux observed from each spot, $ F_i(\phi_i+\phi_0,i,\theta_i,h)$ ($i=1,2$) at a given binary inclination $i$ and colatitude $\theta_i$ of the spot, and the resulting total flux $F_{\rm tot}(\phi)=N(F_1+F_2)$. Here, $N$ is a normalization constant, $\phi_0$ is the spin phase of the primary spot at the folding reference epoch. We assumed the phase of the secondary spot to be $\phi_2=\phi_1+\pi$ and used the relativistic rotating vector model of \citet{poutanen20} to express the phase-resolved 
Stokes parameters of each of the two spots (Eqs.~58--59), assuming that $N F_i$ gives the observed flux and that their polarization degree $p_1=p_2$ does not depend on energy and zenith angle. We then used (Eqs.~21--23) of the same paper to compute the total expected Stokes vector, $Q=Q_1+Q_2$ and $U=U_1+U_2$.

We fitted the phase-resolved Stokes parameters observed before the amplitude increase  (see Fig.~\ref{fig:stokes_intA}) with our simple model, fixing the NS mass and radius to $M=1.4\, \mathrm{M_\odot}$ and $R_{\rm eq}=10$~km  as the fit was insensitive to the choice of these parameters within a range of reasonable values for a NS. The fit (see Fig.~\ref{fig:stokes_intA}) suggests a configuration with two visible spots as the fit $\chi^2$ (38.8 for 41 d.o.f) improves compared to a fit with only one spot ($\chi^2=63.0$ for 42 d.o.f.; the probability of chance improvement is $\simeq 10^{-5}$ according to an F-test). A reasonably good fit can be obtained for any value of $h$ ranging between $0$ and $-1$; here we show the results obtained for a randomly chosen intermediate value of  $h=-0.3$. We find a best-fit for $i=(74.1_{-6.3}^{+5.8})\degr$, $\theta_1=(11.8_{-3.5}^{+2.5})\degr$, $\theta_2=(172.6_{-1.0}^{+2.0})\degr$, $\phi_0=0.57(4)$, $N=6.90(6)\times10^{4}$~cnt, $p_1=4.0\%\pm2.0\%$ and $\chi_{\rm p}=57\fdg2\pm0\fdg5$, where the latter is the position angle of
the pulsar angular momentum. We evaluated uncertainties at 1$\sigma$ confidence level from the variation of the fit $\chi^2$ for the number of interesting parameters measured \citep[$\Delta\chi^2=8.15$ for 7 parameters][]{lampton76}. Such uncertainties have to be taken with caution, as they were obtained for fixed values of the NS mass, radius and emission anisotropy parameter. In addition, the large correlation between the parameters warrants a study of the posterior distribution of the parameters fitted using the likelihood function for $I$, $Q$ and $U$, which will be presented in a forthcoming paper. The red line in Fig.~\ref{fig:stokes_intA} marks the predicted model of the normalized Stokes parameters obtained for these parameters.  Bottom panels of the same figure show the total PD$=\sqrt{q^2+u^2}$ and PA$=\chi_{\rm p}+\frac{1}{2} \tan^{-1}(u/q)$, respectively. The absence of significant variability with the pulse phase of the phase-resolved observed normalized Stokes parameters $q$ and $u$ can be explained by two almost antipodal spots if their magnetic inclination is small. Yet, the observed roughly constant phase-resolved trend gives little information on the other parameters which are essentially set by the total pulse profile alone. \citet{molkov24} used a modeling similar to ours (even though they assumed extended spots) and found qualitatively similar results ($i=58\degr$ and $\theta_1=14\degr$, $\theta_2=\pi-\theta_1$). They modeled a pulse with a higher amplitude (9--12\%) than ours ($A_1=5.2$\%$\pm0.1$\%) and a plateau close to the pulse minimum, which they interpreted in terms of occultation by the inner accretion disk extending down to $\sim$25~km. The flattening of the pulse profile observed by {\ixpe} is less pronounced and partial occultation is not required to obtain qualitative modeling of the profile. Because the pulse profile \citet{molkov24} modeled was observed when the X-ray flux was roughly double that observed by {\ixpe}, this could indicate a recession of the inner accretion disk at a decreasing mass accretion rate.

After MJD  60376.5, the pulsed fraction observed by {\ixpe} increased to $A_1=18.7$\%$\pm0.8$\% (see Fig.~\ref{fig:ixpetiming}). The same model described above with the binary inclination and the anisotropy parameter fixed to $i=74\fdg1$ and $h=-0.3$, respectively, provides a good fit of the profile observed after that epoch, with a spot magnetic inclination and longitude increased by $\simeq 10\degr$ and $\sim$20\degr, respectively, and a polarization degree decreased to $2.0$\%$\pm 1.5$\%, compared to the observations performed previously. The observed change in the pulse profile shape could then reflect motions of the hot spots, as often assumed to explain the timing noise observed from AMSPs \citep[see, e.g., ][]{lamb2009}.
The profile predicted using our model is plotted as a red line in Fig.~\ref{fig:stokes_intB}.

\section{Type-I X-ray bursts}
\label{sec:bursts}

\begin{figure}
\includegraphics[width=8.5cm]{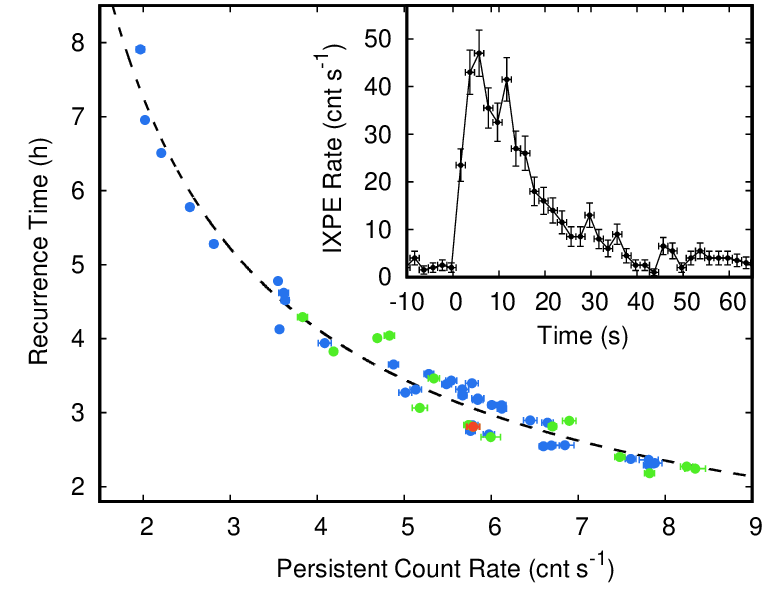}
\caption{Recurrence time of the type-I X-ray bursts observed by {\ixpe} evaluated as the interval between consecutive bursts scaled by a factor 1 (blue points), 2 (green) and 3 (red) as a function of the persistent count rate observed before the burst $C$. The dashed line shows the relation $\Delta t_{\rm rec} = K C^{-0.8}$ with $K=12.75$~h. The inset shows the profile of the burst observed starting at MJD 60377.05969. \label{fig:ixpeburst}}
\end{figure}

During the observations reported here, {\ixpe} detected 52 type-I X-ray bursts which shared similar properties (see the inset of Fig.~\ref{fig:ixpeburst} for an example). The sample of peak count rates had an average of 38.7\,cnt\,s$^{-1}$ and a standard deviation of 2.9\,cnt\,s$^{-1}$. After subtracting the persistent emission, the distribution of the fluence of the  observed bursts had an average of $649$ cnt with a standard deviation of $40$; the fluence of the different bursts was very similar and only slightly increased for bursts occurring when the persistent count rate had decreased below 3\,cnt\,s$^{-1}$. The relatively low number of counts observed by {\ixpe} prevented us from performing a time-resolved spectral analysis of the single bursts. By stacking all the observed bursts, and fitting the observed spectra with the sum of a thermal component and the two-component model used to fit the persistent spectrum (see Sect.~\ref{sec:polanalysis}), we get $\langle kT_{\rm burst}\rangle=1.83\pm0.05$~keV and $R_{\rm  burst}=(6.2\pm0.2) d_8$~km. Defining the burst duration as $\tau=\mathcal{F}/F_{\rm peak}$, and neglecting variations of the spectrum throughout the burst,  gives $ \tau = 16.8\pm1.6$~s. Figure~\ref{fig:ixpeburst} shows the recurrence times $\Delta t_{\rm rec}$ between the observed bursts\footnote{Consecutive bursts without data gaps could not be observed in IXPE data, but the uninterrupted XMM coverage allowed us to define the recurrence time unambiguously.} as a function of the average persistent count rate $C$ measured during the longest possible contiguous interval before the burst. For 14 bursts, we scaled the time elapsed since the previous burst by a factor of 2 as the immediately preceding burst was likely missed due to interruption in the {\ixpe} coverage (green points in Fig.~\ref{fig:ixpeburst}). For one burst (red point), we used a scaling factor of 3.

\subsection{X-ray polarimetry of burst emission}
\label{sec:burstpol}

To search for a polarized signal during the X-ray bursts, we stacked the emission observed during all the events, retaining 32~s after the onset of each of the bursts for a total exposure of 1664~s. We measured $q_{\rm b}=Q_{\rm b}/I_{\rm b}=-0.8\%\pm3.3\%$ and $u_{\rm b}=U_{\rm b}/I_{\rm b}=2.8\%\pm3.3\%$, resulting in an upper limit on the PD of the sum of the burst and the persistent emission of $PD_{\rm{UL}}=7.2\%$ ($90\%$ c.l.).
We then subtracted the absolute Stokes vectors observed during the persistent emission after normalizing them to the exposure time of 1664~s over which the burst emission was integrated, summing in quadrature the uncertainties, and obtained normalized and {\it background} subtracted Stokes parameters $\tilde{q}_{\rm b}=-0.7\% \pm 3.9\%$ and $\tilde{u}_{\rm b} = 3.0\%\pm3.9\%$, corresponding to an upper limit on the PD of 8.5\% (90\% c.l.).

\subsection{Implications}
\label{disc:burst}
To our knowledge, the observations presented here are the first in which {\ixpe} caught thermonuclear X-ray bursts from an accreting NS. 
The $\sim$10~d coverage of the outburst of \mbox{SRGA J1444} allowed us to detect 52 bursts and stack them when searching for a polarized signal. The upper limit we put on the PD of the  burst emission (8.5\%) is higher than the maximum value estimated by \citet{lapidus85} assuming reflection of the burst emission by the accretion disk (3.7\% for an inclination of 72\degr). With a recurrence time variable between $\simeq 1.6$ hr at the beginning of the outburst \citep{molkov24} and $\simeq 8$ hr at the end of the IXPE observation presented here, \mbox{SRGA J1444} was immediately recognized as a very prolific burster. The 11~Hz pulsar IGR J17480--2446  (which showed a recurrence time decreasing from $\simeq 30$ to $\simeq 3$ minutes; \citealt{motta11,chakraborty11,linares12}) and 4U 1636--536 (with a recurrence time of $\simeq 30$ minutes and even less energetic short recurrence time bursts; \citealt{beri19}) are similar cases. Among accreting MSPs, SAX~J17498--2021 showed bursts between 1 and 2 hr \citep{li18}, IGR~J17511--3057 between 7 and 21 hr \citep{falanga11} and IGR~J17498--2921  every 16--18~hr \citep{falanga12}, in all cases with a recurrence time increasing as the X-ray flux declined. The $\approx 15$~s duration of the bursts seen from SRGA J1444 suggests that they ignite in a mixed H/He environment. Hot CNO cycle is expected to deplete the accreted hydrogen in $9.8 (X/0.7) (Z/0.02)^{-1}$~hr, where $H$ and $Z$ are the hydrogen and metals abundances \citep[see][and references therein]{galloway22}.  The range of observed recurrence times ($<8$~hr) suggests that complete depletion of hydrogen could only be achieved if the H abundance is sub-solar. For bursts igniting in a mixed H/He environment, the burst ignition depth is not expected to depend much on the mass accretion rate $\dot{M}$ and the recurrence time scale should increase as $\Delta t \propto \dot{M}^{-1}$. Assuming that the IXPE count rate is a good tracer of the mass accretion rate, the observed dependence (see Fig.~\ref{fig:ixpeburst}) is broadly compatible with such an expectation. Malacaria et al. (in prep.) will present a detailed analysis of the burst ignition process on this fast rotating pulsar allowed by the long and high duty cycle allowed by IXPE.

\section{Conclusions}
\label{conclusions}
We presented the first X-ray polarimetric IXPE observation of an accreting millisecond pulsar in outburst, SRGA J1444, in the context of an observational campaign involving also NICER, NuSTAR and XMM-Newton observations. The main results are the following:

\begin{itemize}

\item The 2--8~keV emission is significantly polarized, with an average degree of $2.3\% \pm 0.4\%$ at an angle of  $59\degr\pm6\degr$ (East of North). 

\item We observed a significant change of the polarization properties with energies. The PD is maximum between 3 and 6~keV ($4.0\%\pm0.5\%$) and decreases to $<2\%$  (90\% c.l.) in the 2--3 keV range.

\item The pulse phase of SRGA J1444 is rather stable throughout the observations presented here, with an upper limit on the spin frequency derivative of $\dot{\nu}<1.2\times10^{-13}$~Hz~s$^{-1}$. During IXPE observations the pulse has an amplitude of $\sim 5$\%, before abruptly increasing to almost $20$\%, roughly one day before the end of the coverage.

\item The observed phase-resolved Stokes parameters are compatible with a constant function. A simple approximate model of the Stokes vector expected from the sum of two almost antipodal spots on the NS surface reproduces the observed values for a binary inclination of $i=(74.1^{+5.8}_{-6.3})\degr$ and relatively small magnetic inclination of the spots $\theta_1=(11.8_{-3.5}^{+2.5})\degr$, $\theta_2=(172.6_{-1.0}^{+2.0})\degr$. The little phase-resolved variability shown by the Stokes parameters, if any, prevented us from obtaining constraints on the NS mass and radius based on our simple modeling. 

\item IXPE observed 52 type-I X-ray bursts sharing similar properties (e.g. burst peak fluence, duration, fluence) with a recurrence time increasing as a function of the observed count rate, $\Delta t_{\rm rec}\propto C^{-0.8}$. After subtracting the persistent emission, the upper limit on the PD observed in the stacked burst emission is  $8.5$\% (90\% c.l.).

\end{itemize}

The first significant detection of polarization from an accreting millisecond pulsar confirms theoretical expectations and strengthens the prospects of using this technique to measure the geometrical parameters of accreting MSPs and obtain constraints on the mass and radius of the NS from pulse profile modeling. Future IXPE observations of accreting millisecond pulsars outbursts will hopefully detect a larger phase-resolved variability and enable more accurate determination than those allowed by the present dataset.

\begin{acknowledgements}
We warmly thank the Directors and the Science Operations Team of IXPE, NICER, NuSTAR, and XMM for promptly scheduling the observations reported here and Bas Dorsman, Matteo Bachetti and Anna Watts for useful discussions..

The Imaging X-ray Polarimetry Explorer (IXPE) is a joint US and Italian mission.  The US contribution is supported by the National Aeronautics and Space Administration (NASA) and led and managed by its Marshall Space Flight Center (MSFC), with industry partner Ball Aerospace (contract NNM15AA18C).  The Italian contribution is supported by the Italian Space Agency (Agenzia Spaziale Italiana, ASI) through contract ASI-OHBI-2022-13-I.0, agreements ASI-INAF-2022-19-HH.0 and ASI-INFN-2017.13-H0, and its Space Science Data Center (SSDC) with agreements ASI-INAF-2022-14-HH.0 and ASI-INFN 2021-43-HH.0, and by the Istituto Nazionale di Astrofisica (INAF) and the Istituto Nazionale di Fisica Nucleare (INFN) in Italy.  This research used data products provided by the IXPE Team (MSFC, SSDC, INAF, and INFN) and distributed with additional software tools by the High-Energy Astrophysics Science Archive Research Center (HEASARC), at NASA Goddard Space Flight Center (GSFC).  
NICER is a 0.2--12\,keV X-ray telescope operating on the International Space Station, funded by NASA. 
The NuSTAR mission is a project led by the California Institute of Technology, managed by the Jet Propulsion Laboratory, and funded by the National Aeronautics and Space Administration. Data analysis was performed using the NuSTAR Data Analysis Software (NuSTARDAS), jointly developed by the ASI Science Data Center (SSDC, Italy) and the California Institute of Technology (USA). 
XMM-Newton is an ESA science mission with instruments and contributions directly funded by ESA Member States and NASA.

MN and this work were supported by NASA under grant 80NSSC24K1175. AP, GI, FA, RLP, CM and LS are supported by INAF (Research Grant ''Uncovering the optical beat of the fastest magnetised neutron stars (FANS)'') and the Italian Ministry of University and Research (MUR) (PRIN 2020, Grant 2020BRP57Z, ''Gravitational and Electromagnetic-wave Sources in the Universe with current and next-generation detectors (GEMS)''). AP acknowledges support from the Fondazione Cariplo/Cassa Depositi e Prestiti, grant no. 2023-2560. JP thanks the Ministry of Science and Higher Education  grant 075-15-2024-647 for support. 
AB acknowledges support from the Finnish Cultural Foundation grant 00240328. 
MCB acknowledges support from the INAF-Astrofit fellowship. 
FCZ is supported by a Ram\'on y Cajal fellowship (grant agreement RYC2021-030888-I).
TS acknowledges support from ERC Consolidator Grant No.~865768 AEONS (PI: Watts).
\end{acknowledgements}

\bibliographystyle{aa}
\bibliography{main.bib}

\end{document}